\documentclass[%
reprint,
superscriptaddress,
showpacs,preprintnumbers,
 amsmath,amssymb,
 aps,
 longbibliography,
 prb,
floatfix,
]{revtex4-1}

\usepackage{epsfig}
\usepackage[thinspace]{SIunits}
\usepackage{xspace}
\usepackage{graphicx}
\graphicspath{{./}}
\usepackage{dcolumn}
\usepackage{bm}
\usepackage{color}
\usepackage[normalem]{ulem}
\usepackage{hyperref}
\hypersetup{
    unicode=false,
    pdftoolbar=true,
    pdfmenubar=true,
    pdffitwindow=false,
    pdfstartview={FitH},
    pdftitle={My title},
    pdfauthor={Author},
    pdfsubject={Subject},
    pdfcreator={Creator},
    pdfproducer={Producer},
    pdfkeywords={keyword1} {key2} {key3},
    pdfnewwindow=true,
    colorlinks=true,
    linkcolor=blue,
    citecolor=blue,
    filecolor=blue,
    urlcolor=blue,
    plainpages=false
}
\usepackage[symbol*]{footmisc}
\DefineFNsymbolsTM{otherfnsymbols}{%
	\textdagger	\dagger
	\textsection	\mathsection
	\textasteriskcentered	*
  \textdaggerdbl	\ddagger
  \textbardbl	\|%
  \textparagraph	\mathparagraph
	\textbullet	\circ
}%
\setfnsymbol{otherfnsymbols}

\newcommand{\Tc}{\ensuremath{T_\mathrm{c}}\xspace}
\newcommand{\Ts}{\ensuremath{T_\mathrm{s}}\xspace}
\newcommand{\BFA}{\mbox{BaFe$_{2}$As$_2$}\xspace}
\newcommand{\FeSe}{\mbox{FeSe}\xspace}
\newcommand{\Alg}{\texorpdfstring{\ensuremath{A_{1g}}\xspace}{A1g}}
\newcommand{\AZg}{\texorpdfstring{\ensuremath{A_{2g}}\xspace}{A2g}}
\newcommand{\Blg}{\texorpdfstring{\ensuremath{B_{1g}}\xspace}{B1g}}
\newcommand{\BZg}{\texorpdfstring{\ensuremath{B_{2g}}\xspace}{B2g}}
\newcommand{\wn}{\ensuremath{\rm cm^{-1}}\xspace}

\begin{document}
\title{Frustrated spin order and stripe fluctuations in FeSe}
\date{\today}
\author{A. Baum}
\affiliation{Walther Meissner Institut, Bayerische Akademie der Wissenschaften, 85748 Garching, Germany}
\affiliation{Fakult\"at f\"ur Physik E23, Technische Universit\"at M\"unchen, 85748 Garching, Germany}
\author{H. N. Ruiz}
\affiliation{Stanford Institute for Materials and Energy Sciences, SLAC National Accelerator Laboratory, 2575 Sand Hill Road, Menlo Park, California 94025, USA}
\affiliation{Department of Physics, Stanford University, California 94305, USA}
\author{N. Lazarevi\'c}
\affiliation{Center for Solid State Physics and New Materials, Institute of Physics Belgrade, University of Belgrade, Pregrevica 118, 11080 Belgrade, Serbia}
\author{Yao Wang}
\affiliation{Stanford Institute for Materials and Energy Sciences, SLAC National Accelerator Laboratory, 2575 Sand Hill Road, Menlo Park, California 94025, USA}
\affiliation{Department of Applied Physics, Stanford University, California 94305, USA}
\author{T.~B\"ohm}
\altaffiliation{Present address: TNG Technology Consulting GmbH, Beta-Stra\ss{}e, 85774 Unterf\"{o}hring, Germany}
\affiliation{Walther Meissner Institut, Bayerische Akademie der Wissenschaften, 85748 Garching, Germany}
\affiliation{Fakult\"at f\"ur Physik E23, Technische Universit\"at M\"unchen, 85748 Garching, Germany}
\author{R.~Hosseinian~Ahangharnejhad}
\altaffiliation{Present address: School of Solar and Advanced Renewable Energy, Department of Physics and Astronomy, University of Toledo, Toledo, Ohio 43606, USA}
\affiliation{Walther Meissner Institut, Bayerische Akademie der Wissenschaften, 85748 Garching, Germany}
\affiliation{Fakult\"at f\"ur Physik E23, Technische Universit\"at M\"unchen, 85748 Garching, Germany}
\author{P. Adelmann}
\affiliation{Karlsruher Institut f\"ur Technologie, Institut f\"ur Festk\"orperphysik, 76021 Karlsruhe, Germany}
\author{T. Wolf}
\affiliation{Karlsruher Institut f\"ur Technologie, Institut f\"ur Festk\"orperphysik, 76021 Karlsruhe, Germany}
\author{Z.\,V. Popovi\'c}
\affiliation{Center for Solid State Physics and New Materials, Institute of Physics Belgrade, University of Belgrade, Pregrevica 118, 11080 Belgrade, Serbia}
\affiliation{Serbian Academy of Sciences and Arts, Knez Mihailova 35, 11000 Belgrade, Serbia}
\author{B.~Moritz}
\affiliation{Stanford Institute for Materials and Energy Sciences, SLAC National Accelerator Laboratory, 2575 Sand Hill Road, Menlo Park, California 94025, USA}
\author{T.\,P. Devereaux}
\affiliation{Stanford Institute for Materials and Energy Sciences, SLAC National Accelerator Laboratory, 2575 Sand Hill Road, Menlo Park, California 94025, USA}
\affiliation{Geballe Laboratory for Advanced Materials, Stanford University, California 94305, USA}
\author{R. Hackl}
\email{hackl@wmi.badw.de}
\affiliation{Walther Meissner Institut, Bayerische Akademie der Wissenschaften, 85748 Garching, Germany}

\begin{abstract}
  The charge and spin dynamics of the structurally simplest iron-based superconductor, FeSe, may hold the key to understanding the physics of high temperature superconductors in general. Unlike the iron pnictides, FeSe lacks long range magnetic order in spite of a similar structural transition around 90\,K. Here, we report results of Raman scattering experiments as a  function of temperature and polarization and simulations based on exact diagonalization of a frustrated spin model. Both experiment and theory find a persistent low energy peak close to 500\,cm$^{-1}$ in $B_{1g}$ symmetry, which softens slightly around 100\,K, that we assign to spin excitations. By comparing with results from neutron scattering, this study provides evidence for nearly frustrated stripe order in FeSe.
\end{abstract}
\maketitle


\section*{Introduction}
Fe-based pnictides and chalcogenides, similar to cuprates, manganites or some heavy fermion compounds, are characterized by the proximity and competition of various phases including magnetism, charge order and superconductivity. Specifically the magnetism of Fe-based systems has various puzzling aspects which do not straight\-forwardly follow from the Fe valence or changes in the Fermi surface topology \cite{Yin:2011,Georges:2013,Si:2016,Skornyakov:2017}. Some systems have a nearly ordered localized moment close to $2\,\mu{\rm _B}$ \cite{LiSL:2009a}, such as FeTe or rare-earth iron selenides, whereas the moments of $A$Fe$_2$As$_2$-based compounds ($A=$\,Ba, Sr, Eu or Ca) are slightly below $1\,\mu_{\rm B}$ \cite{Johnston:2010} and display aspects of itinerant spin-density-wave (SDW) magnetism with a gap in the electronic excitation spectrum \cite{YiM:2017}.  In contrast others do not order down to the lowest temperatures, such as FeSe \cite{Baek:2014} or LaFePO \cite{Taylor:2013}.

The material specific differences are a matter of intense discussion, and low- as well as high-energy electronic and structural properties determine the properties \cite{Mazin:2009,Yin:2011,Georges:2013,Stadler:2015,Glasbrenner:2015,Skornyakov:2017,Baum:2018a}. At the Fermi energy $E_{\rm F}$, the main fraction of the electronic density of states originates from $t_{2g}$ Fe orbitals, but a substantial part of the Fe-Fe hopping occurs via the pnictogen or chalcogen atoms,  hence via the $xz$, $yz$, and $p_z$ orbitals. For geometrical reasons, the resulting exchange coupling energies between nearest ($J_1$) and next nearest neighbour ($J_2$) iron atoms have the same order of magnitude, and small changes in the pnictogen (chalcogen) height above the Fe plane influence the ratio $J_2/J_1${, such that} various {orders} are energetically very close \cite{Glasbrenner:2015}.

The reduced overlap of the in-plane $xy$ orbitals decreases the hopping integral $t$  and increases the influence of the Hund's rule interactions and the correlation energy $U$, even though they are only in the range of 1-2\,eV. Thus the electrons in the $xy$ orbitals have a considerably higher effective mass $m^\ast$ and smaller quasiparticle weight $Z$ than those of the $xz$ and $yz$ orbitals. This effect was coined orbital selective Mottness \cite{Anisimov:2002,deMedici:2009,deMedici:2017} and observed by photoemission spectroscopy (ARPES) in Fe-based chalcogenides \cite{YiM:2015}. It is similar in spirit to what was found by Raman scattering in the cuprates as a function of momentum \cite{Venturini:2002b}. In either case some of the electron wave functions are more localized than others. This paradigm may explain why the description remains difficult and controversial in all cases.

Therefore we address the question as to whether systematic trends can be found across the families of the Fe-based {superconductors}, how the spin excitations are related to other highly correlated systems, and how they can be described appropriately.

As an experimental tool we use Raman scattering since the differences expected theoretically \cite{Yin:2011,Si:2016} and indicated experimentally in the electronic structure \cite{YiM:2017} can be tracked in both the charge and the spin channel. Another advantage is the large energy range of approximately 1\,meV to 1\,eV (8 to 8,000\,\wn) accessible by light scattering \cite{Devereaux:2007}.

Early theoretical work on Fe-based systems considered the Heisenberg model the most appropriate approach \cite{Chen:2011b}, and the high-energy maxima observed by Raman scattering in BaFe$_2$As$_2$ were interpreted in terms of localized spins \cite{Sugai:2011,Sugai:2012}.  On the other hand, the low-energy spectra are reminiscent of charge density wave (CDW) or SDW formation \cite{Chauviere:2011,Sugai:2012,Eiter:2013,YangYX:2014}. In principle, both effects can coexist if the strength of the correlations varies for electrons from different orbitals, where itinerant electrons form a SDW, while those on localized orbitals give rise to a Heisenberg-like response.

In contrast to the $A$Fe$_2$As$_2$-based compounds ($A=$\,Ba, Sr, Ca), FeSe seems to be closer to localized order with a larger mass renormalization than in the iron pnictides \cite{Yin:2011}.
Apart from low lying charge excitations, the remaining, presumably spin, degrees of freedom in FeSe may be adequately described by a spin-1 $J_1$-$J_2$-$J_3$-$K$ Heisenberg model \cite{Glasbrenner:2015}
which provides also a consistent description of our results shown in this work and allows for the presence of different spin orders. Since various types of spin order are energetically in close proximity  \cite{Glasbrenner:2015,Wang:2015,WangQS:2016}, frustration may quench long-range order down to the lowest temperatures \cite{Baek:2014}, even though neutron scattering experiments in FeSe find large values for the exchange energies \cite{Rahn:2015,WangQS:2016}.

Recent experiments on FeSe focused on low-energies and \Blg ($x^2-y^2$) symmetry, and the response was associated with particle-hole excitations and critical fluctuations \cite{Massat:2016}.
Here, we obtain similar experimental results below 1,500\,\wn. Those in the range 50-200\,\wn show similarities with the other Fe-based systems while those above 200\,\wn are distinctly different but display similarities with the cuprates \cite{Sulewski:1991,Muschler:2010a}. In addition to previous work, we analyze all symmetries at {higher} energies up to 3,500\,\wn, to uncover crucial information about the behaviour of the spin degrees of freedom.

By comparing experimental and simulated Raman data we find a persistent low energy peak at roughly 500\,\wn in \Blg symmetry, which softens slightly around 100\,K. We assign the $B_{1g}$ maximum and the related structures in \Alg and \BZg symmetry to spin excitations. The  theoretical simulations also aim at establishing a link between light and neutron scattering data with respect to the spin degrees of freedom and to furnish evidence for nearly frustrated stripe order at low temperature.
We arrive at the conclusion that frustrated order of localized spins dominates the physics in FeSe, while critical spin and/or charge fluctuations are not the main focus of the paper.

\begin{figure*}
  \centering
  \includegraphics[width=17cm]{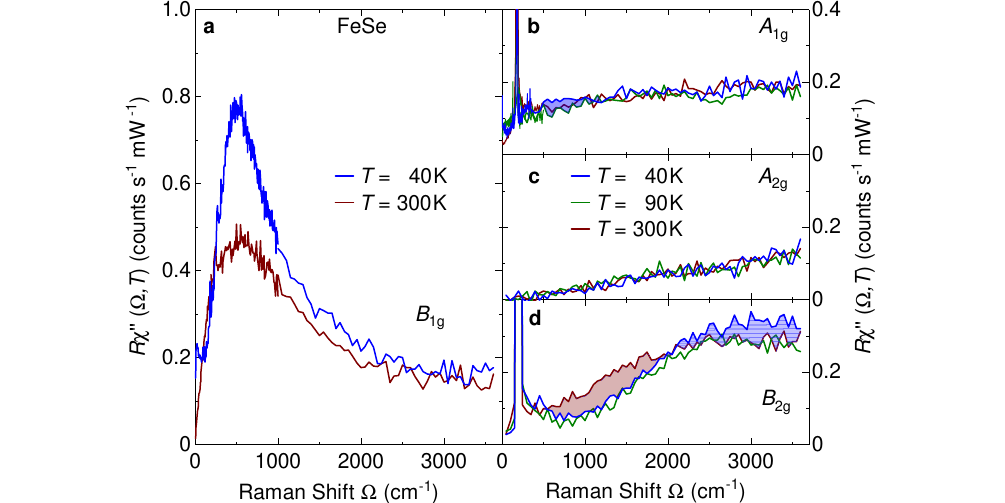}
  \caption{Symmetry-resolved Raman spectra of FeSe at various temperatures for large energy transfers. \textbf{a} \Blg spectra at temperatures as indicated. The spectrum at 90\,K is omitted here for clarity but is displayed in a separate figure below 
  The weak structure at $T=40$\,K in the range 20-25\,\wn is left over from the fluctuation peak which is most pronounced right above \Ts as shown below.
  \textbf{b} \Alg, \textbf{c} \AZg, and \textbf{d} \BZg spectra at temperatures as indicated. In \Alg and \BZg symmetry particle-hole excitations dominate the response. In agreement with the simulations weak additional peaks from spin excitations appear at low temperature (blue shaded areas). \BZg shows a loss of spectral weight (shaded red). The narrow lines close to 200\,cm$^{-1}$ are the \Alg and \Blg phonons. In the 1\,Fe unit cell used here the \Blg phonon appears in \BZg symmetry since the axes are rotated by 45$^\circ$ with respect to the crystallographic (2\,Fe) cell. The \AZg intensity vanishes below 500\,\wn and the cross section is completely temperature independent.
  }
  \label{fig:data-sym-long}
\end{figure*}

\section*{Results}
\label{sec:results}
\subsection*{Experiments}
Symmetry-resolved Raman spectra of single-crystalline FeSe (see Methods) in the energy range up to 0.45\,eV (3,600\,cm$^{-1}$) are shown in Fig.~\ref{fig:data-sym-long}. The spectra are linear combinations of the polarization dependent raw data (see Methods and Supplementary Fig.~1 in Supplementary Note~1). For \Blg symmetry (Fig.~\ref{fig:data-sym-long}a) we plot only two temperatures, 40 and 300\,K, to highlight the persistence of the peak at approximately 500\,cm$^{-1}$. The full temperature dependence will be shown below.
For \Alg, \AZg, and \BZg symmetry we show spectra at 40, 90 and 300\,K (Fig.~\ref{fig:data-sym-long}b-d).
Out of the four symmetries, the \Alg, \Blg, and \BZg spectra display Raman active phonons, magnons or electron-hole excitations, while the \AZg spectra are weak and vanish below 500-1,000\,cm$^{-1}$. As intensity in \AZg symmetry appears only under certain conditions not satisfied in the present study, we ignore it here.

In the high-energy limit the intensities are smaller in all symmetries than those in other Fe-based systems such as BaFe$_2$As$_2$ (see Supplementary Fig.~2 in Supplementary Note~2). However, in the energy range up to approximately 3,000\,cm$^{-1}$  there is a huge additional contribution to the \Blg cross section in FeSe (Fig.~\ref{fig:data-sym-long}a). The response is strongly temperature dependent and peaks at 530\,cm$^{-1}$ in the low-temperature limit.
Between 90 and 40\,K the \Alg and \BZg spectra increase slightly in the range around 700 and 3,000\,cm$^{-1}$, respectively (indicated as blue shaded areas in Fig.~\ref{fig:data-sym-long}b and d). The overall intensity gain in the \Alg and \BZg spectra in the shaded range is a fraction of approximately 5\% of that in \Blg symmetry. The \BZg spectra exhibit a reduction in spectral weight in the range from 600 to 1,900\,\wn (shaded red) which is already fully developed at the structural transition at $T_\mathrm{s} = 89.1\,\mathrm{K}$ in agreement with earlier work \cite{Massat:2016}. In contrast to \Alg and \BZg symmetry, the temperature dependence of the \Blg intensity is strong, whereas the peak energy changes only weakly, displaying some similarity with the cuprates \cite{Knoll:1990}. This similarity, along with the considerations of Glasbrenner \textit{et al.} \cite{Glasbrenner:2015}, motivated us to explore a spin-only, Heisenberg-like model for describing the temperature evolution of the Raman scattering data.

\begin{figure}
  \centering
  \includegraphics[width=8.5cm]{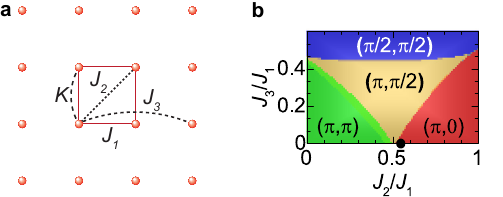}
  \caption{Model and resulting phase diagram. \textbf{a} A $4\times 4$ cluster was used for the simulations. The red spheres represent the Fe atoms, each of which carries localized spin $\mathbf{S}_{i}$, with $S=1$. The nearest, next-nearest, and next-next-nearest neighbour interactions $J_1$, $J_2$, and $J_3$, respectively, are indicated. $K$ is the coefficient of the biquadratic term proportional to $({\bf S}_i \cdot {\bf S}_j)^2$.
  \textbf{b} $J_2-J_3$ phase diagram as obtained from our simulations at $T=0$ and for $K=0.1$. The black dot shows the  parameters at which temperature-dependent simulations have been performed.
  }
  \label{fig:cluster}
\end{figure}

\begin{figure}
  \centering
  \includegraphics[width=8.5cm]{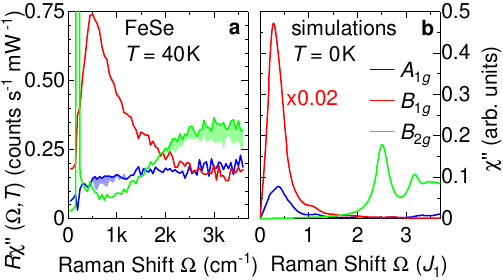}
  \caption{Symmetry-resolved Raman spectra of FeSe for large energy shifts at low temperature. \textbf{a} Experimental results for symmetries as indicated at 40\,K. The \Blg peak at 500\,cm$^{-1}$ dominates the spectrum. In \Alg and \BZg symmetry the electron-hole continua dominate the response, and the magnetic excitations yield only small additional contributions at approximately 700 and 3,000\,cm$^{-1}$, respectively. \textbf{b} Simulated Raman spectra at $T=0\,K$ including only magnetic contributions. The \Alg and \Blg symmetries have peaks solely at low energies whereas the \BZg contributions are at high energies only. The \Blg response is multiplied by a factor of 0.02.
  }
  \label{fig:data-low}
\end{figure}

\subsection*{Simulations at zero temperature}
We performed numerical simulations at zero temperature for a frustrated spin-1 system on the basis of a $J_1$-$J_2$-$J_3$-$K$ Heisenberg model \cite{Glasbrenner:2015} on a  16-points cluster as shown in Fig.~\ref{fig:cluster}a and described in the Methods section. Fig.~\ref{fig:cluster}b shows the resulting phase diagram as a function of $J_2$ and $J_3$. $K$ was set at 0.1 (repulsive) in order to suppress ordering tendencies on the small cluster. The parameter set for the simulations of the Raman and neutron data at finite temperature is indicated as a black dot.

In Fig.~\ref{fig:data-low} we show the low-temperature data (Fig.~\ref{fig:data-low}a) along with the simulations (Fig.~\ref{fig:data-low}b).  The energy scale for the simulations is given in units of $J_1$ which has been derived to be 123\,meV or 990\,cm$^{-1}$  \cite{Glasbrenner:2015}, allowing a semi-quantitative comparison with the experiment. As already mentioned, the experimental \Alg and \BZg spectra are not dominated by spin excitations and we do not attempt to further analyze the continua extending to energies in excess of 1\,eV, considering them a background. The opposite is true for \Blg symmetry, also borne out in the simulations. For the selected values of $J_1=123$\,meV, $J_2=0.528\,J_1$, $J_3=0$, and $K=0.1\,J_1$, the positions of the spin excitations in the three symmetries and the relative intensities are qualitatively reproduced. The choice of parameters is motivated by the previous use of the $J_{1}$-$J_{2}$ Heisenberg model, with $J_{1}=J_{2}$ to describe the stripe phase of iron pnictides \cite{Chen:2011b}. Here we use a value of $J_{2}$ smaller than $J_{1}$ to enhance competition between N\'{e}el and stripe orders when describing FeSe. This approach and choice of parameters is strongly supported in a recent neutron scattering study \cite{WangQS:2016}.

The comparison of the different scattering symmetries, the temperature dependence, and our simulations indicate that the excitation at 500\,cm$^{-1}$ is an additional scattering channel superimposed on the particle-hole continuum and fluctuation response, as shown in Supplementary Note~3 with Supplementary Figures~3 and 4.  Here we focus on the peak centered at approximately 500\,cm$^{-1}$ which, in agreement with the simulations, originates from two-magnon excitations in a highly frustrated spin system, although the features below 500\,cm$^{-1}$ also are interesting and were interpreted in terms of quadrupolar orbital fluctuations \cite{Massat:2016}.

\subsection*{Temperature dependence}
\begin{figure}
  \centering
  \includegraphics[width=8.5cm]{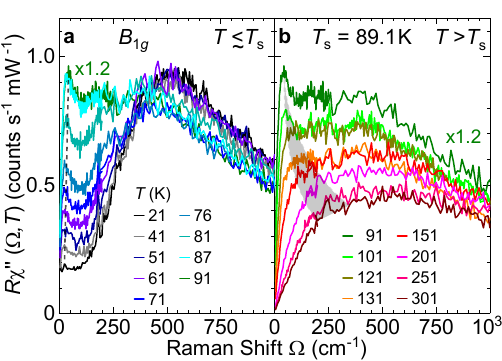}
  \caption[]{Raman spectra in \Blg symmetry 
  at temperatures \textbf{a} below and \textbf{b} above $\Ts = 89.1\,\mathrm{K}$. The spectrum at 91\,K appears in both panels for better comparison. The black dashed line in panel \textbf{a} and the grey shaded area in panel \textbf{b} indicate the approximate positions of the low energy peak resulting from critical fluctuations. The peak centered close to 500\,cm$^{-1}$ results from excitations of neighboring spins which are studied here. A tentative decomposition is shown in Supplementary Fig.~4.
  }
  \label{fig:temp_Blg}
\end{figure}

It is enlightening to look at the \Blg spectra across the whole temperature range as plotted in Fig.~\ref{fig:temp_Blg}. The well-defined two-magnon peak centered at approximately 500\,cm$^{-1}$ in the low temperature limit loses intensity, and becomes less well defined with increasing temperature up to the structural transition $\Ts = 89.1\,\mathrm{K}$.  Above the structural transition, the spectral weight continues to decrease and the width of the two-magnon feature grows, while the peak again becomes well-defined and the energy hardens slightly approaching the high temperature limit of the study.  What may appear as a gap opening at low temperature is presumably just the reduction of spectral weight in a low-energy feature at approximately 22\,cm$^{-1}$.  The intensity of this lower energy response increases with temperature, leading to a well-formed peak at an energy around 50\,cm$^{-1}$ near the structural transition.  Above the structural transition this feature rapidly loses spectral weight, hardens, and becomes indistinguishable from the two-magnon response in the high temperature limit.  This low-energy feature develops in a fashion very similar to that found in Ba(Fe$_{1-x}$Co$_x$)$_2$As$_2$ for $x>0$ \cite{Choi:2008,Gallais:2013,Kretzschmar:2016}.

\begin{figure}
  \centering
  \includegraphics[width=8.5cm]{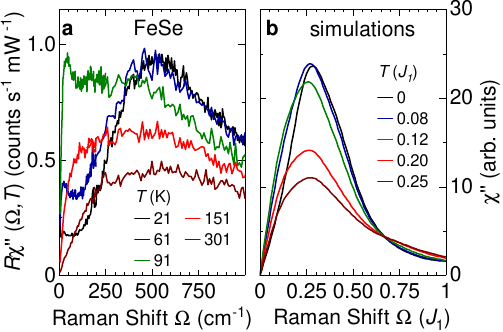}
  \caption{Temperature dependence of the \Blg response. \textbf{a}~Experimental spectra at selected temperatures as indicated. The spectra include several excitations the decomposition of which is shown in Supplementary Fig.~4. \textbf{b} Simulated Raman response at temperatures as indicated. Only magnetic excitations are included. The coupling constant was derived as $J_1 = 123$\,meV in Ref.~\onlinecite{Glasbrenner:2015}.
  }
  \label{fig:B1g-T}
\end{figure}

Now we compare the measurements with numerical simulations for the temperature dependence of the Raman \Blg susceptibility in Fig.~\ref{fig:B1g-T}a and b, respectively. For the simulations (Fig.~\ref{fig:B1g-T}b) we use the same parameters as at $T=0$ (black dot in Fig.~\ref{fig:data-low}). At zero temperature the simulations show a single low energy \Blg peak around 0.3\,$J_{1}$. As temperature increases, a weak shoulder forms on the low energy side of the peak, and the whole peak softens slightly and broadens over the simulated temperature range.
Except for the additional intensity at low energies, $\Omega < 200\,\wn$, (Fig.~\ref{fig:B1g-T}a) there is good qualitative agreement between theory and experiment.
As shown in Supplementary Fig.~5 in Supplementary Note~4, a similar agreement between experiment and simulations is obtained for the temperature dependence in \Alg and \BZg symmetries, indicating that both the gain in intensity (blue shaded areas in Fig.~\ref{fig:data-sym-long}) as well as the reduction in spectral weight in \BZg from 600 to 1,900\,\wn (shaded red in Fig.~\ref{fig:data-sym-long}d) can be attributed to the frustrated localized magnetism.

\subsection*{Connection to the spin structure factor}
\begin{figure}
  \centering
  \includegraphics[width=8.5cm]{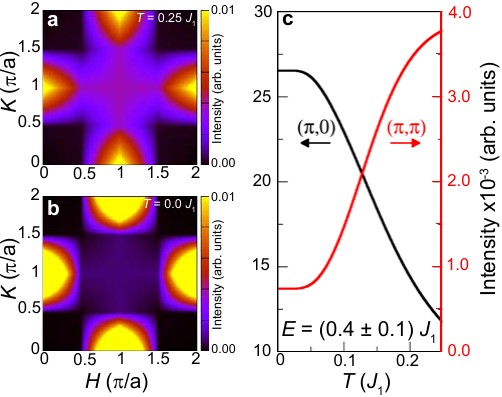}
  \caption{Simulations of the dynamical structure factor $S({\bf q},\omega)$ of localized spin excitations integrated over an energy window of $(0.4\pm 0.1)\,J_1$. \textbf{a} and \textbf{b} display cuts through the first Brillouin zone at $T=0.25$ and $0\,J_1$, respectively. At high temperature there is intensity at $(\pi,\pi)$ indicating a tendency towards N\'eel order. At low temperature the intensity at $(\pi,\pi)$ is reduced and the stripe-like antiferromagnetism with $(\pi,0)$ ordering wave vector becomes stronger.  \textbf{c} {$S({\bf q},\omega)$} integrated over an energy window $(0.4\pm 0.1)\,J_1$ for fixed momenta $(\pi,\pi)$ and $(\pi,0)$ {intensities} as a function of temperature.
  }
  \label{fig:neutrons}
\end{figure}

To support our explanation of the Raman data, we simulated the dynamical spin structure factor $S({\bf q},\omega)$ and compared the findings to results of neutron scattering experiments \cite{WangQS:2016}. While clearly not observing long-range order, above the structural transition neutron scattering finds similar intensity at finite energy for several wave vectors along the line $(\pi,0)-(\pi,\pi)$. Upon cooling, the spectral weight at these wave vectors shifts away from $(\pi,\pi)$ to directions along $(\pi,0)$, although the respective peaks remain relatively broad. In Fig.~\ref{fig:neutrons}a and b we show the results of the simulations for two characteristic temperatures. As temperature decreases, spectral weight shifts from $(\pi, \pi)$ towards $(\pi,0)$ in agreement with the experiment \cite{WangQS:2016}. In Fig.~\ref{fig:neutrons}c we show the evolution of the spectral weights around $(\pi,\pi)$ and $(\pi,0)$ in an energy window of $(0.4\pm0.1)\,J_1$ as a function of temperature, similar to the results shown in Ref.~\onlinecite{WangQS:2016}.  In the experiment, the temperature where the integrated dynamical spin structure factor changes most dramatically is close to the structural transition. From our simulations, the temperature where similar changes occur in comparison to neutron scattering corresponds to the temperature at which the simulated \Blg response (Fig.~\ref{fig:B1g-T}) shows the most pronounced shoulder, and the overall intensity begins to decrease. Not surprisingly, the low-energy peak in the Raman scattering experiment is also strongest near the structural transition.

\section*{Discussion}
The agreement of experiment with theory in both neutron and Raman scattering suggests that a dominant contribution to the FeSe spectra comes from frustrated magnetism of essentially local spins. The differences between the classes of ferro-pnictides and -chalcogenides, in particular the different degrees of itineracy,  may then originate in a subtle orbital differentiation across the families \cite{Yin:2011}.

\begin{figure}
  \centering
  \includegraphics[width=8.5cm]{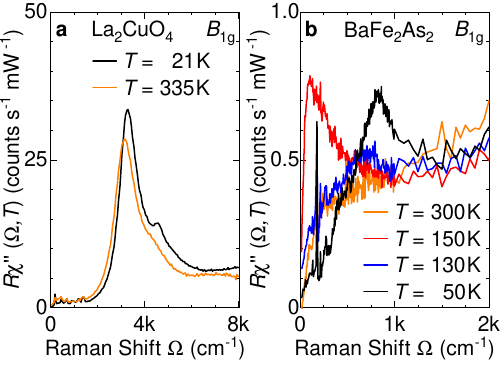}
  \caption{Examples of localized and itinerant magnets. \textbf{a} \Blg Raman spectra of La$_2$CuO$_{4}$. From Ref. \onlinecite{Muschler:2010a}. \textbf{b} \Blg spectra of BaFe$_2$As$_2$ at four characteristic temperatures as indicated.
  }
  \label{fig:LCO-Ba122}
\end{figure}

If FeSe were frustrated, near such a phase boundary between magnetic states, then its behaviour would be consistent with the observed sensitivity to intercalation \cite{Burrard:2013,ZhangA:2013}, layer thickness \cite{GeJF:2015}, and pressure \cite{Medvedev:2009}, which could affect the exchange interactions through the hopping.  Relative to the theoretical results below 200\,cm$^{-1}$, critical fluctuations of any origin, which are characterized by a diverging correlation length close to the transition, can neither be described nor distiguished in such a small cluster calculation. Here, only experimental arguments can be applied similar to those in Ref.~\onlinecite{Kretzschmar:2016}, but will not be further discussed, since they are not the primary focus of the analysis. A brief summary may be found in the Supplementary Note~3.

It is remarkable how clearly the Raman spectra of an SDW state originating from a Fermi surface instability and a magnet with local moments can be distingiuished. For comparison, Fig.~\ref{fig:LCO-Ba122} shows \Blg Raman spectra for La$_2$CuO$_4$ and BaFe$_2$As$_2$ at characteristic temperatures. La$_2$CuO$_4$ (Fig.~\ref{fig:LCO-Ba122}a) is an example of a material with local moments on the Cu sites \cite{Sulewski:1991,Muschler:2010a} having a N\'eel temperature of $T_{\rm N}=325$\,K. The well-defined peak at approximately 2.84\,$J_1$ \cite{Canali:1992,Weidinger:2015} possesses a weak and continuous temperature dependence across $T_{\rm N}$ \cite{Knoll:1990}. The origin of the scattering in La$_2$CuO$_4$ and other insulating cuprates \cite{Chelwani:2018} can thus be traced back to Heisenberg-type physics of local moments \cite{Fleury:1968}, which, for simplicity, need only include the nearest-neighbour exchange interaction $J_1$.

In contrast, most iron-based superconductors are metallic antiferromagnets in the parent state exhibiting rather different Raman signatures.  In BaFe$_2$As$_2$ (Fig.~\ref{fig:LCO-Ba122}b) abrupt changes are observed in \Blg symmetry upon entering the SDW state: the fluctuation peak below 100\,cm$^{-1}$ vanishes, a gap develops below some 500-600\,cm$^{-1}$, and intensity piles up in the range 600-1,500\,cm$^{-1}$ \cite{Chauviere:2010,Sugai:2012}, the typical behaviour of an SDW or CDW \cite{Eiter:2013} in weak-coupling, resulting from Fermi surface nesting. Yet, even for itinerant systems such as these, longer range exchange interactions can become relevant and lead to magnetic frustration \cite{Yildirim:2009b}.

In summary, the Raman response of FeSe was measured in all symmetries and compared to simulations of a frustrated spin-1 system. The experimental data were decomposed in order to determine which parts of the spectra originate from particle-hole excitations, fluctuations of local spins, and low energy critical fluctuations.
Comparison of the decomposed experimental data with the simulations gives evidence that the dominant contribution of the Raman spectra comes from magnetic competition between $(\pi,0)$ and $(\pi, \pi)$ ordering vectors. These features of the Raman spectra, which agree qualitatively with a spin only model, consist of a dominant peak in \Blg symmetry around 500\,\wn along with a peak at similar energy but lower intensity in \Alg and at higher energy in \BZg symmetry. These results will likely help to unravel the mechanism behind the superconducting phase found in FeSe.


\section*{Methods}
\subsection*{Experiment}

The FeSe crystals were prepared by the vapor transport technique. Details of the
crystal growth and characterization are described elsewhere
\cite{Bohmer:2013}. Before the experiment the samples were cleaved in air and the exposure time was minimized. The surfaces obtained in this way have several atomically flat regions allowing us to measure spectra down to 5\,cm$^{-1}$. At the tetragonal-to-orthorhombic transition $T_{\rm s}$ twin boundaries appear and become clearly visible in the observation optics. As described in detail by Kretzschmar \textit{et al.} \cite{Kretzschmar:2016} the appearance of stripes can be used to determine the laser heating $\Delta T_{\rm L}$ and $T_{\rm s}$ to be $(0.5\pm0.1)\,\mathrm{K}\,\mathrm{mW}^{-1}$ and $(89.1\pm0.2)$\,K, respectively.

Calibrated Raman scattering equipment was used for the experiment. The samples were attached to the cold finger of a He-flow cryostat having a vacuum of approximately $5\cdot10^{-5}$\,Pa ($5\cdot10^{-7}$\,mbar). For excitation we used a diode-pumped solid state laser emitting at 575\,nm (Coherent GENESIS MX-SLM 577-500) and various lines of an Ar ion laser (Coherent Innova 304). The angle of incidence was close to $66^\circ$ for reducing the elastic stray light entering the collection optics. Polarization and power of the incoming light were adjusted in a way that the light inside the sample had the proper polarization state and, respectively, a power of typically $P_a=4$\,mW independent of polarization. For the symmetry assignment we use the 1\,Fe unit cell (axes $x$ and $y$ parallel to the Fe-Fe bonds) which has the same orientation as the magnetic unit cell in the cases of N\'eel or single-stripe order (4\,Fe cell). The orthorhombic distortion is along these axes whereas the crystallographic cell assumes a diamond shape with the length of the tetragonal axes preserved. Because of the rotated axes in the 1\,Fe unit cell the Fe $B_{1g}$ phonon appears in the $B_{2g}$ spectra.
Spectra at low to medium energies were measured with a resolution $\sigma \approx 5\,\wn$ in steps of $\Delta \Omega = 2.5$ or 5\,\wn below 250\,\wn and steps of 10\,\wn above where no sharp peaks need to be resolved. Spectra covering the energy range up to 0.5-1\, eV were measured with a resolution $\sigma \approx 20\,\wn$ in steps of $\Delta \Omega = 50\,\wn.$

\subsection*{Simulations}
We use exact diagonalization to study a Heisenberg-like model on a 16 site square lattice, which contains the necessary momentum points and is small enough that exact diagonalization can reach high enough temperatures to find agreement with the temperature dependence in the experiment. This was solved using the parallel Arnoldi method \cite{Sorensen:1998}. The Hamiltonian is given by
\begin{equation}
\begin{split}
\mathcal{H}=\sum_{\rm nn}[J_{1} \mathbf{S}_{i} \cdot \mathbf{S}_{j} + K(\mathbf{S}_{i} \cdot \mathbf{S}_{j})^{2}]  \\ + \sum_{\rm 2nn} J_{2} \mathbf{S}_{i} \cdot \mathbf{S}_{j} + \sum_{\rm 3nn} J_{3} \mathbf{S}_{i} \cdot \mathbf{S}_{j}
\vspace{2mm}
\end{split}
\end{equation}
where $\mathbf{S}_{i}$ is a spin-1 operator reflecting the observation that the local moments of iron chalcogenides close to 2\,$\mu_\mathrm{B}$ \cite{Gretarsson:2011}. The sum over nn is over nearest neighbours, the sum over 2nn is over next nearest neighbours, and the sum over 3nn is over next next nearest neighbours.

We determine the dominant order according to the largest static spin structure factor, given by
\begin{equation} \label{eq:StaticStructureFactor}
S(\mathbf{q}) = \frac{1}{N} \sum_{l} e^{i\mathbf{q} \cdot \mathbf{R}_{l}} \sum_{i}  \langle{\mathbf{S}_{\mathbf{R}_{i}+\mathbf{R}_{l}} \cdot \mathbf{S}_{\mathbf{R}_{i}}}\rangle.
\end{equation}
Due to the possible spontaneous symmetry breaking we adjust the structure factor by the degeneracy of the momentum. To characterize the relative strength of the dominant fluctuations we project the relative intensity of the dominant static structure factor onto the range [0,1] using the following
\begin{equation} \label{eq:intensity}
  \text{intensity} = 1 - \frac {d_{\mathbf{q}_\mathrm{sub}}S(\mathbf{q}_\mathrm{sub})}{d_{\mathbf{q}_\mathrm{max}}
  S(\mathbf{q}_\mathrm{max})}
\end{equation}
where $d_{\mathbf{q}}$ is the degeneracy of momentum $\mathbf{q}$, $\mathbf{q}_{\rm max}$ is the momentum with the largest $d_{\mathbf{q}}S_{\mathbf{q}}$, and $\mathbf{q}_{\rm sub}$ is the momentum with the second largest (subdominant) $d_{\mathbf{q}}S_{\mathbf{q}}$.

The Raman susceptibilities for \Blg, \BZg, and \Alg symmetries for non-zero temperatures were calculated using the Fleury-Loudon scattering operator \cite{Chen:2011b} given by

\begin{equation}
\mathcal{O}=\sum_{i,j} J_{ij} (\hat{\mathbf{e}}_\mathrm{in} \cdot \hat{\mathbf{d}}_{ij}) (\hat{\mathbf{e}}_\mathrm{out} \cdot \hat{\mathbf{d}}_{ij}) \mathbf{S}_{i} \cdot \mathbf{S}_{j}
\end{equation}
\noindent
where $J_{ij}$ are the exchange interaction values used in the Hamiltonian, $\hat{\mathbf{d}}_{ij}$ is a unit vector connecting sites $i$ and $j$ and $\hat{\mathbf{e}}_\mathrm{in/out}$ are the polarization vectors. For the symmetries calculated we use the polarization vectors
\begin{equation}
\begin{gathered}
\hat{\mathbf{e}}_\mathrm{in}\!=\!\frac{1}{\sqrt{2}}(\hat{\mathbf{x}}+\hat{\mathbf{y}}),~ \hat{\mathbf{e}}_\mathrm{out}\!=\!\frac{1}{\sqrt{2}}(\hat{\mathbf{x}}+\hat{\mathbf{y}}) \textrm{ for } \Alg \oplus \BZg,\\
\hat{\mathbf{e}}_\mathrm{in}=\hat{\mathbf{x}},~ \hat{\mathbf{e}}_\mathrm{out}=\hat{\mathbf{y}} \textrm{ for } \BZg,\\
\hat{\mathbf{e}}_\mathrm{in}=\frac{1}{\sqrt{2}}(\hat{\mathbf{x}}+\hat{\mathbf{y}}),~ \hat{\mathbf{e}}_\mathrm{out}=\frac{1}{\sqrt{2}}(\hat{\mathbf{x}}-\hat{\mathbf{y}}) \textrm{ for } \Blg,
\end{gathered}
\end{equation}
(where $\hat{\mathbf{x}}$ and $\hat{\mathbf{y}}$ point along the Fe-Fe directions). We use this operator to calculate the Raman response $R(\omega)$ using the continued fraction expansion \cite{Dagotto:1994}, where $R(\omega)$ is given by

\begin{widetext}
\begin{equation} \label{eq:R}
R(\omega) = -\frac{1}{\pi Z} \sum_{n} e^{-\beta E_{n}} \operatorname{Im}\left(\langle{\Psi_{n}}|{\mathcal{O^{\dagger}} \frac{1}{\omega + E_{n} + i\epsilon - \mathcal{H}} \mathcal{O}}|{\Psi_{n}}\rangle\right)
\end{equation}

\end{widetext}
\noindent
with $Z$ the partition function. The sum traverses over all eigenstates $\Psi_{n}$ of the Hamiltonian $\mathcal{H}$ having eigenenergies $E_n < E_{0}+2J_{1}$ where $E_{0}$ is the ground state energy. The Raman susceptibility is given by $\chi^{\prime\prime}(\omega) = \frac{1}{2}[R(\omega) - R(-\omega)]$. The dynamical spin structure factor was calculated using the same method with $\mathcal{O}$ replaced with $S_{\mathbf{q}}^{z}=\frac{1}{\sqrt{N}}\sum_{l}e^{i\mathbf{q} \cdot \mathbf{R}_{l}} S_{l}^{z}$.

\section*{Acknowledgement}
The work was supported by the German Research Foundation (DFG) via the Priority
Program SPP\,1458 (grant-no. Ha2071/7) and the Transregional Collaborative Research Center TRR80 and by the Serbian Ministry of Education, Science and Technological Development under Project III45018. We acknowledge support by the DAAD through the bilateral project between Serbia and Germany (grant numbers 57142964 and 57335339). The collaboration with Stanford University was supported by the Bavaria California Technology Center BaCaTeC (grant-no. A5\,[2012-2]). Work in the SIMES at Stanford University and SLAC was supported by the U.S. Department of Energy, Office of Basic Energy Sciences, Division of Materials Sciences and Engineering, under Contract No. DE-AC02-76SF00515. Computational work was performed using the resources of the National Energy Research Scientific Computing Center supported by the U.S. Department of Energy, Office of Science, under Contract No. DE-AC02-05CH11231.

\section*{Author contributions}

A.B., T.B., and R.H. conceived the experiment.
B.M. and T.P.D. conceived the ED analysis.
P.A. and T.W. synthesized and characterized the samples.
A.B., N.L., T.B., and R.H.A. performed the Raman scattering experiment.
H.N.R. and Y.W. coded and performed the ED calculations.
A.B., H.N.R., N.L., B.M., and R.H. analyzed and discussed the data.
A.B., H.N.R., N.L., Z.P., B.M., T.P.D., and R.H. wrote the paper.
All authors commented on the manuscript.

\section*{Competing interests}
The authors declare that there are no competing interests.

\section*{Data availability}
Data are available upon reasonable request from the corresponding author.
~~~~~~~~~

\renewcommand*{\figurename}{Supplementary~Figure}
    \makeatletter
    \renewcommand*{\fnum@figure}{{\figurename~\thefigure}}
    \renewcommand*{\@caption@fignum@sep}{{. }}
    \makeatother
\setcounter{figure}{0}
\setcounter{equation}{0}

\section*{Supplementary Note 1: Polarization dependence of the Raman spectra of \texorpdfstring{\FeSe}{FeSe}}
\label{Asec:polarization}

\begin{figure}
  \centering
  \includegraphics[width=8.5cm]{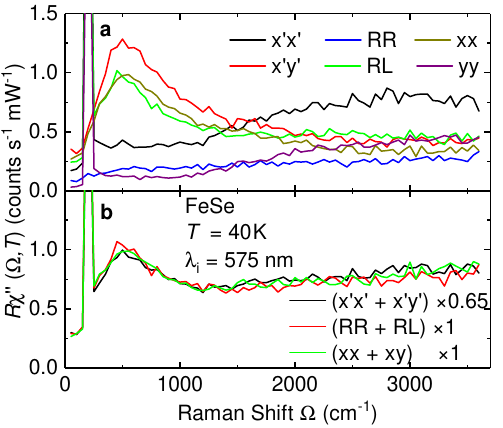}
  \caption[]{
  Raman spectra of FeSe at $T=40\,K$. \textbf{a} Spectra at polarizations in the FeSe plane as indicated. The $x$ and $y$ axes run along the Fe-Fe bonds and  $x^\prime$ and $y^\prime$ are rotated by 45$^\circ$. \textbf{b} Sums of complementary spectra each yielding the full set of all four accessible symmetries. The spectra are multiplied as indicated.}
  \label{Afig:data-pol-long}
\end{figure}

Supplementary Fig.~\ref{Afig:data-pol-long}a shows the complete set of polarization resolved Raman spectra we measured for FeSe at $T=40$\,K up to a maximum energy of $0.45$\,eV. The measured spectra have been corrected for the sensitivity of the instrument and divided by the Bose factor $\left\{ 1-\exp(-\frac{\hbar\Omega}{k_\mathrm{B} T})\right\}$. In Supplementary Fig.~\ref{Afig:data-pol-long}b sums of corresponding pairs of spectra are shown. Each sum contains the full set of all four symmetries ($\Alg + \AZg + \Blg + \BZg$) accessible with the light polarizations in the Fe plane. All three sets exhibit the same spectral shape. The spectra measured with linear light polarizations at 45$^\circ$ with respect to the Fe-Fe bonds ($x^\prime x^\prime$ and $x^\prime y^\prime$) were multiplied by a factor of 0.65 to fit the other configurations. The same factor was applied when calculating the sums for extracting the pure symmetries. The reason for this deviation from the expected $x^\prime x^\prime$ and $x^\prime y^\prime$ intensities lies in small inaccuracies in determining the optical constants. Since we never observed polarization leakages the main effect pertains obviously on the power absorption and transmission rather than phase shifts between the parallel and perpendicular light polarizations.

\section*{Supplementary Note 2: Raman spectra of \texorpdfstring{\BFA}{BaFe2As2}}
\label{Asec:BFA}
\begin{figure}
  \centering
  \includegraphics[width=8.5cm]{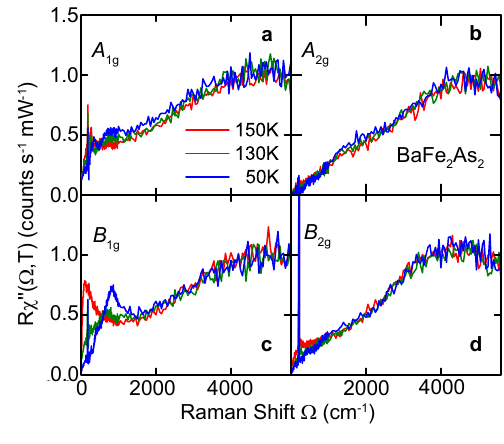}
  \caption[]{Symmetry-resolved spectra of \BFA for three temperatures $150\,K > T_\mathrm{SDW}$, $130\,K \lesssim T_\mathrm{SDW}$ and $50\,K \ll T_\mathrm{SDW}$.
  }
  \label{Afig:BFA}
\end{figure}

Supplementary Fig.~\ref{Afig:BFA} shows the Raman spectra of BaFe$_2$As$_2$ as a function of symmetry and temperature. Towards high energies the spectra increase almost monotonically over an energy range of approximately 0.7\,eV. We could not observe the pronounced nearly polarization-independent maxima in the range 2,000 - 3,000\,cm$^{-1}$ reported in Ref.~\onlinecite{Sugai:2010}. At high energies our spectra are temperature independent. At low energies pronounced changes are observed in $A_{1g}$ and $B_{1g}$ symmetry upon entering the striped spin density wave (SDW) state below $T_{\rm SDW} = 135$\,K as described by various authors \cite{Chauviere:2010,Chauviere:2011,Sugai:2012}. In $A_{2g}$ and $B_{2g}$ symmetry the changes are small but probably significant in that polarization leakage is unlikely to be the reason for the weak low-temperature peaks in the range 2,000\,cm$^{-1}$ and the gap-like behaviour below approximately 1,000\,cm$^{-1}$. The changes are particularly pronounced in $B_{1g}$ symmetry. As shown in Supplementary Fig.~\ref{Afig:BFA}c, in Fig.~1b of the main text and in more detail elsewhere \cite{Kretzschmar:2016} the fluctuation peak vanishes very rapidly and the redistribution of spectral weight from low to high energies sets in instantaneously at $T_{\rm SDW}$. All these observations show that the polarization and temperature dependences here are fundamentally different from those of FeSe (Fig.~1 of the main text).

\section*{Supplementary Note 3: Delineation of the contributions to the \Blg spectra}
\label{Asec:AL-fit}
\begin{figure}
  \centering
  \includegraphics[width=8.5cm]{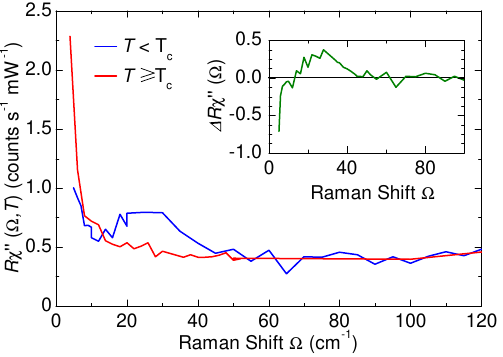}
  \caption[]{Raman spectra of FeSe in \Blg symmetry above (red) and below (blue) the superconducting transition at $T_c = 8.8$\,K. The inset shows the difference between the two spectra $R\Delta\chi^{\prime\prime}(\Omega) = R\chi^{\prime\prime}(\Omega,\,T<T_c) - R\chi^{\prime\prime}(\Omega,T \geq T_c)$.}
  \label{Afig:SC}
\end{figure}

Supplementary Fig.~\ref{Afig:SC} shows Raman spectra of the FeSe sample at temperatures below (blue line) and above (red line) the superconducting transition temperature \Tc, which was determined to be $\Tc = 8.8$\,K by measuring the third harmonic of the magnetic susceptibility \cite{Venturini:2002d}. Both spectra show a sharp increase towards the laser line which can be attributed to increased elastic scattering due to an accumulation of surface layers at low temperatures. Below \Tc a broad peak emerges centred around approximately 28\,cm$^{-1}$ which we identify as pair breaking peak at $2\Delta \approx (4.5\pm0.5)\,k_\mathrm{B} \Tc$. Above 50\,cm$^{-1}$ the spectra at $T<\Tc$ and $T \geq \Tc$ are identical. We could not resolve the second peak close to 40\,cm$^{-1}$ as observed in Ref.~\onlinecite{Massat:2016}. The gap ratio of $(4.5\pm0.5)\, k_\mathrm{B} \Tc$ is comparable to what was found for Ba(Fe$_{0.939}$Co$_{0.061}$)$_2$As$_2$ \cite{Muschler:2009} but smaller than that found for Ba$_{1-x}$K$_x$Fe$_2$As$_2$ \cite{Bohm:2017}.
The existence of a superconducting gap and a pair-breaking peak in the Raman spectra shows that the magnetic features are superposed on an electronic continuum.

%
The temperature and symmetry dependence of the  Raman response (Figs.~1 and 4 of the main text) indicate that the \Blg spectra are a superposition of various scattering channels as shown in Supplementary Fig.~\ref{Afig:AL-fit}: (i) particle-hole excitations and presumably also a weak contribution from luminescence in the range up to 1\,eV and beyond, (ii) critical fluctuations of either spin or charge in the range below 250\,cm$^{-1}$, and (iii) excitations of neighboring spins with the response centered at 500\,cm$^{-1}$ in \Alg and \Blg symmetry and at 3,000\,cm$^{-1}$ in \BZg symmetry.

(i) An estimate of electron-hole excitations may be obtained by comparing the \Blg with the \Alg and \BZg spectra at various temperatures including $T<T_c$. In a first approximation we assume that luminescence has a weak symmetry and temperature dependence and find that the intensities in all channels have the same order of magnitude. We use the \BZg continuum for deriving an analytical approximation for modeling the particle-hole spectrum (blue in Supplementary Fig.~\ref{Afig:AL-fit}).

\begin{figure*}
  \centering
  \includegraphics[width=15cm]{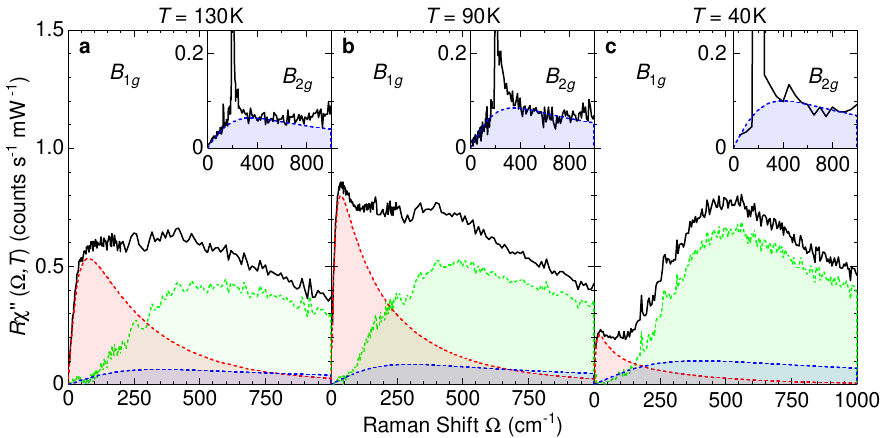}
  \caption{Phenomenological description of the data at three characteristic temperatures. The experimental Raman spectra (black) are a superposition of the particle-hole continuum (blue), here estimated from the \BZg spectra (inset), critical fluctuations (red)  \cite{Caprara:2005}, and excitations of neighboring spins (green).%
  }
  \label{Afig:AL-fit}
\end{figure*}

(ii) There are various ways to derive the Raman response of critical fluctuations with finite wave vector ${\bf Q}_c$. Caprara and coworkers considered the clean limit and, consequently, calculated the response and the selection rules for a pair of fluctuations having ${\bf Q}_c$ and $-{\bf Q}_c$ thus maintaining the ${\bf q}=0$ selection rule for light scattering \cite{Caprara:2005}. Alternatively the the collision-limited regime was considered where the momentum of the fluctuation can be carried away by an impurity \cite{Gallais:2016a}. Finally, quadrupolar fluctuations in the unit cell can give rise to Raman scattering \cite{Thorsmolle:2016,Massat:2016}. In either case the response diverges at or slightly below the structural transition where the correlation length diverges. We used the approach of Ref.~\onlinecite{Caprara:2005} for modeling the response since we believe that FeSe is in the clean limit and that spin fluctuations are a possible candidate for the response \cite{Kretzschmar:2016}. Yet, the decision about the type of fluctuations relevant here is not a subject of this publication, and we are predominantly interested in excitations of neighboring spins.

(iii) For isolating the response of neighboring spins in the total Raman response we subtract the particle-hole continuum (i) and the response of fluctuations (ii) from the spectra. The resulting difference is shown in green in Supplementary Fig.~\ref{Afig:AL-fit} and can be considered the best possible approximation to the two-magnon response. At temperatures much smaller or larger than \Ts the critical fluctuations do not contribute substantially to the total response and can be ignored. The particle-hole continuum is generally weak. Therefore the simulations can be best compared to the Raman data at temperatures sufficiently far away from \Ts as shown in Figs.~3 and 5. Since the simulations were performed on a $4\times4$ cluster critical fluctuations cannot be described close to \Ts where the correlation length is much larger than the cluster size.

\section*{Supplementary Note 4: Temperature dependence in \texorpdfstring{\Alg}{A1g} and \texorpdfstring{\BZg}{B2g} symmetries}
\label{Asec:temp_symmetries}
\begin{figure}
  \centering
  \includegraphics[width=8.5cm]{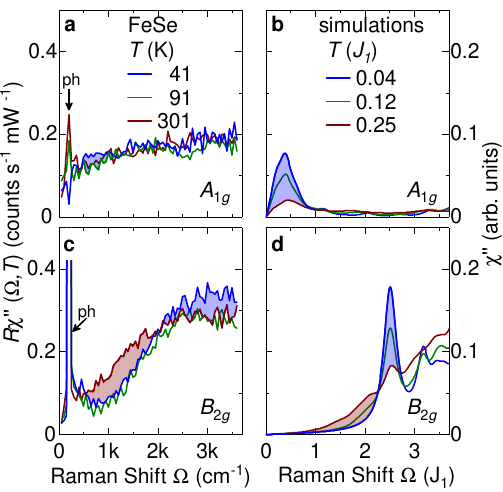}
  \caption[]{Simulations at $T>0$ for \Alg and \BZg symmetry. \textbf{a},\textbf{c} Experimental and \textbf{b},\textbf{d} simulated Raman spectra of FeSe in \textbf{a},\textbf{b} \Alg and \textbf{c},\textbf{d} \BZg symmetries for temperatures as indicated. Loss and gain of spectral weight upon cooling are indicated by the red and blue shaded areas, respectively. $J_1 \approx 123\,\mathrm{meV}\,\hat{=}\,990\,\mathrm{cm}^{-1}$ \cite{Glasbrenner:2015}. The sharp peaks in the experimental spectra marked with ``ph'' are phonons.%
	}
  \label{Afig:temp_sym}
\end{figure}

Supplementary Figure~\ref{Afig:temp_sym} compares experimental and simulated Raman spectra in \Alg and \BZg symmetry up to high energies at room temperature (red), slightly above \Ts (green) and below \Ts (blue). The choice of temperatures for the simulated spectra corresponds to Fig.~5 of the main text. Sharp phonon peaks (labelled ph) appear in the experimental spectra at 200\,cm$^{-1}$ the shape of which is not reproduced properly since resolution and sampling width are reduced. With $J_1 \approx 123\,\mathrm{meV}$ (990\,cm$^{-1}$) as found in Ref.~\onlinecite{Glasbrenner:2015} the experimental and simulated spectra can be compared semi-quantitatively. Both theory and experiment consistently show a gain in intensity for \Alg at medium energies and for \BZg at high energies (blue shaded areas in the respective spectra) as well as a reduction of spectral weight in \BZg in the range from 600 to 1,900\,\wn (shaded red). The changes appear to be more continuous in the simulations than in the experiment where the gain in intensity in both symmetries only occurs at $T<\Ts$. The reduction in spectral weight in \BZg symmetry has already taken place at \Ts (green spectra).

%

\end{document}